\documentclass{webofc}

\usepackage[varg]{txfonts}   
\usepackage{hyperref}
\usepackage{url}
\hypersetup{colorlinks=true,citecolor=blue,urlcolor=blue,linkcolor=blue}
\usepackage{microtype}
\begin{document}
\title{Hard probes and nuclear PDFs}
\author{\firstname{Petja} \lastname{Paakkinen}\inst{1,2}\fnsep\thanks{\email{petja.k.m.paakkinen@jyu.fi}}}
\institute{University of Jyväskylä, P.O. Box 35, FI-40014 University of Jyväskylä, Finland \and Helsinki Institute of Physics, P.O. Box 64, FI-00014 University of Helsinki, Finland}
\abstract{A mini-review on the status of global analyses of nuclear parton distribution functions is given, focusing on the most relevant constraints for the hard-probes phenomenology in ultra-relativistic heavy-ion collisions.}
\maketitle
\section{Introduction}
\label{intro}
In ultra-relativistic heavy-ion collisions (HICs), the production of jets and heavy quarks---i.e.\ \emph{hard probes}---plays a special role. Due to their large transverse momentum ($p_{\rm T}$) or mass, these probes are formed at the early stages of HICs through hard scatterings of high-energy partons (quarks and gluons). The production mechanism is therefore understood in terms of perturbative QCD (pQCD), and their later interactions with the produced bulk matter can be used in verifying the creation of a hot QCD phase in such collisions and determining its properties~\cite{Akiba:2015vaa,Apolinario:2022vzg,Pablos:2024yxw}. However, in order to have these probes well calibrated, one needs to know the distributions of the initial-state partons in the colliding nuclei precisely enough. This requires the study of nuclear parton distribution functions (nuclear PDFs, or nPDFs for short). The dependence of hard-probe production on nPDFs has been demonstrated multiple times, ranging from the suppression of very-high-$p_{\rm T}$ jets~\cite{Caucal:2020uic} to comparisons with early-stage glasma effects in heavy-quark production at low $p_{\rm T}$~\cite{Avramescu:2024poa}. Recently, the issues with large uncertainties in the initial-state nuclear modifications for the search of hot-QCD energy-loss effects in light-ion collisions have also been raised~\cite{Brewer:2021tyv,Paakkinen:2021jjp,Gebhard:2024flv}.

In the global-analysis approach, nPDFs are extracted from fits to inclusive hard cross section data. The idea is to use lepton--nucleus ($l+A$) and hadron--nucleus ($h+A$) data to avoid complications from hot-QCD effects present in larger collision systems. For purely electroweak (EW) observables, where the probe does not interact with the created medium, or in the case of photoproduction processes in ultraperipheral collisions (UPCs), where the nuclei pass each other without direct hadronic interaction and no hot matter is created, it is possible to use also nucleus--nucleus ($A+A$) collision data to constrain the nPDFs. In the absence of hot-QCD effects, one can calculate the relevant observables purely within the QCD collinear factorisation, and then use statistical inference to fit model-agnostic parametrisations of nPDFs to the available data. The working assumption in these fits is thus that the fluidlike collective effects in small systems~\cite{Grosse-Oetringhaus:2024bwr} and additional cold nuclear matter effects (see e.g.\ Ref.~\cite{Arleo:2021bpv}) beyond nuclear modifications of the PDFs are relevant only in rare high-multiplicity events or at low $p_{\rm T}$, therefore not influencing the hard inclusive observables (integrated over final-state particle configurations and multiplicities) at high enough $p_{\rm T}$ in the $p+A$ and reference proton--proton ($p+p$) collisions. With new data becoming available, these assumptions and their region of validity can be systematically tested.

This mini-review works as an overview of the most constraining processes in nPDF fits relevant for the phenomenology of the hard and EW probes in HICs. A more comprehensive review with extensive references to the datasets used in global analyses and discussion on the physics-origin of the nuclear modifications can be found in Ref.~\cite{Klasen:2023uqj}.

\section{Probes with leptonic final states}
\label{sec-leptonic}
\begin{figure}[htb]
\centering
\includegraphics[width=13.2cm]{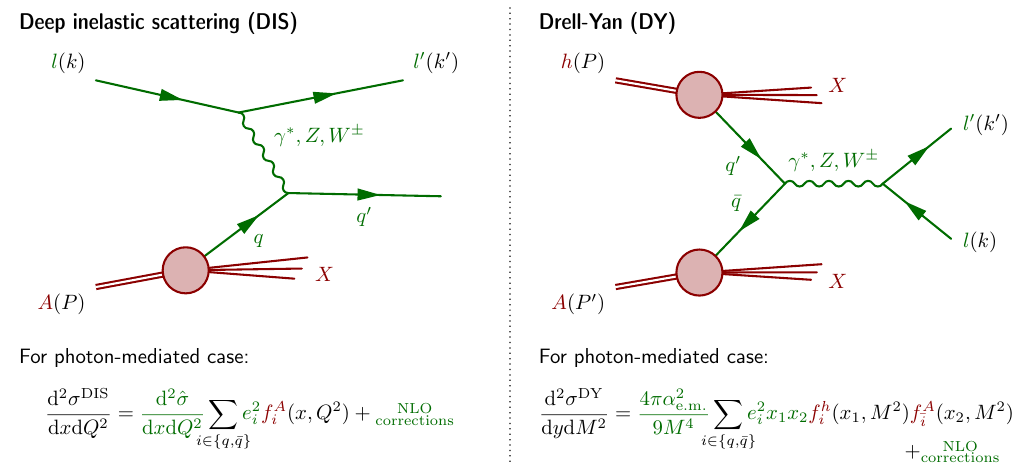}
\caption{Illustrations and pQCD expressions for the DIS (left) and DY (right) processes.}
\label{fig-leptonic}
\end{figure}

The backbone process for any nPDF analysis is that of deep inelastic scattering (DIS), $l+A \rightarrow l'+X$, with $X$ denoting ``anything'', illustrated in figure~\ref{fig-leptonic} (left). For the virtual-photon mediated case this process factorises, at leading order (LO) in pQCD, into a product of the partonic cross section $\hat{\sigma}$ and a charge-squared-weighted sum of the quark PDFs $f_i^A(x, Q^2)$, $i \in \{q, \bar{q}\}$, where the Bjorken variable $x = Q^2 / 2P \cdot (k - k')$ and virtuality $Q^2 = - (k - k')^2$ that are accessed through external kinematics, enter directly as the longitudinal momentum fraction of the parton and the probing scale for the PDFs. Accounting for next-to-leading order (NLO) corrections turns this product into a convolution over the momentum fraction and gives rise to a contribution from the gluon PDF. While much of the available fixed-target data are rather old by now and the constraints are well established, the impact from neutrino-induced DIS is still subject to active research~\cite{Muzakka:2022wey,Helenius:2024fow}, and new data from large-$x$ charged-lepton-DIS measurements at JLab are being included in the global analyses~\cite{Paukkunen:2020rnb,Segarra:2020gtj}. The Electron Ion Collider~\cite{Accardi:2012qut} will revolutionarise the study of nuclear DIS with access to smaller $x$ than ever before.

Further insight on the nPDFs can be derived from the Drell--Yan (DY) process $h+A \rightarrow ll'+X$, shown in figure~\ref{fig-leptonic} (right), producing a pair of leptons with the invariant mass $M^2 = (k + k')^2$, or the EW-boson mass, providing a hard scale for the process. Again, gluon PDFs enter only at NLO, and the DIS and DY processes are therefore essential in fixing the quark nPDFs. Here, the CERN-LHC provides an access to the ``standard candle'' EW-boson production processes in $p+A$ and $A+A$ collisions (see Ref.~\cite{Klasen:2023uqj} for references to available data).
With two hadrons in the initial state, the DY process involves a product (at LO) or convolution (at NLO) of two PDFs at different momentum fractions $x_{1,2}$. This poses a problem for extracting the nPDFs from $p+A$ DY data, as the observed cross section now depends on the PDFs of the projectile proton, inducing a source of uncertainty~\cite{Paukkunen:2010qg,Eskola:2022rlm}. The approaches taken in nPDF analyses vary. While TUJU21~\cite{Helenius:2021tof}, nCTEQ15HQ~\cite{Duwentaster:2022kpv} and nNNPDF3.0~\cite{AbdulKhalek:2022fyi} use absolute $p+{\rm Pb}$ cross sections, EPPS21~\cite{Eskola:2021nhw} uses $p+{\rm Pb}/p+p$ nuclear modification ratios when possible to cancel dependence on the free-proton PDFs. In some cases, like for the $W^\pm$-boson charge asymmetry, the sensitivity to proton PDFs can be particularly important~\cite{Paukkunen:2010qg}. Whether the large difference between the Run~2 data and predictions in a forward (proton-going) direction in $p+{\rm Pb}$ observed at ALICE~\cite{ALICE:2022cxs} is due to nuclear or proton PDFs thus requires further study. Alternative asymmetry observables with reduced proton-PDF dependence due to utilising also the rapidity asymmetry of the collision system have also been proposed~\cite{Paukkunen:2010qg,Eskola:2022rlm}.
For the $Z/\gamma^*$-mediated dilepton production, the TUJU21 analysis~\cite{Helenius:2021tof} demonstrated that to properly describe the normalisation of the data in the low-invariant-mass bin of the CMS Run~2 $p+{\rm Pb}$ measurement~\cite{CMS:2021ynu}, it was necessary to include the next-to-next-to-leading order (NNLO) corrections for this process---a first clear evidence for the need of NNLO-accurate nPDFs.

\section{Hadroproduction of hadronic final states}
\label{sec-hadronic}

\begin{figure}[htb]
\centering
\includegraphics[width=13.2cm]{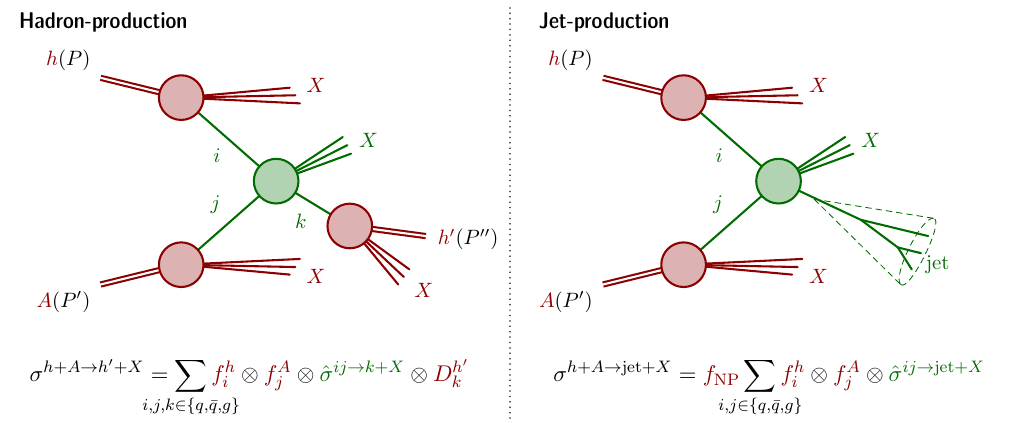}
\caption{Illustrations and pQCD expressions for the inclusive hadron (left) and jet (right) hadroproduction processes.}
\label{fig-hadronic}
\end{figure}

Since the gluon PDF enters in the DIS and DY processes only at NLO, processes with hadronic final states, where the gluons are involved already at LO, have proven indispensible for constraining the nPDFs. The price to pay is that hadronisation dynamics has to be accounted for in a way or another. For the inclusive hadron production $h+A \to h'+X$, shown in figure~\ref{fig-hadronic} (left), one accounts for the hadronisation effects with the parton-to-hadron fragmentation functions $D_k^{h'}(z,Q^2)$ with a convolution over the momentum fraction $z$, which come as a source of uncertainty for the nPDF fits. A lot of this uncertainty cancels in the $p+A/p+p$ ratios, if the hadronisation is assumed not to be significantly modified in the $p+A$ collisions compared to $p+p$, and good fits to available data have been reached in analyses including both light~\cite{Duwentaster:2021ioo} and heavy flavour (HF) data~\cite{Duwentaster:2022kpv,AbdulKhalek:2022fyi,Eskola:2021nhw}. Very recently, however, the LHCb $p+{\rm Pb}/p+p$ ratio measurement for $\pi^0$ production~\cite{LHCb:2022tjh} showed an enhancement in the backward (lead-going) direction compared to nPDFs predictions in the $2\ {\rm GeV} < p_{\rm T} < 4\ {\rm GeV}$ range, while in the forward direction these data are compatible with constraints from D$^0$ production data~\cite{LHCb:2017yua}. Similar enhancement in this $p_{\rm T}$ range might be present also in the ALICE midrapidity $\pi^0$ data~\cite{ALICE:2021est}. If this enhancement persists in closer inspections, this could mean that the traditional $p_{\rm T} > 3\ {\rm GeV}$ cut for hadron production in is too small and needs to be re-addressed in future nPDF fits. Furthermore, in the case of HF production, which currently gives the strongest constraints for small-$x$ gluons, different pQCD mass schemes, Monte Carlo event generator implementations or data-driven surrogate models have been used in the fits (see the discussion in Ref.~\cite{Paakkinen:2022qxn}), with corresponding impact on the extracted nPDFs.

Complementary to inclusive hadron production, one can use the inclusive jet production in $h+A$ collisions, see figure~\ref{fig-hadronic} (right), for studying the nPDFs at large momentum-trasfer scales. Now, instead of fragmentation functions, one needs an infrared-safe definition of a jet, and since only the production of partonic jets is calculated in fixed-order pQCD, one has to apply additional non-perturbative (NP) corrections $f_{\rm NP}$ to take into account hadronisation and underlying-event multi-parton-interaction effects that can modify the jet kinematics. For the correct interpretation of the data, it is therefore important that experimental analyses document whether, and how, possible underlying-event subtractions are done. Currently, the CMS measurement of $p+{\rm Pb}/p+p$ nuclear modification ratio of dijet self-normalised spectra~\cite{CMS:2018jpl} have been used as a constraint in the EPPS21~\cite{Eskola:2021nhw} and nNNPDF3.0~\cite{AbdulKhalek:2022fyi} analyses, finding reduction in the gluon nPDF uncertainties. Regarding this dataset, it is still unclear why these fits fail to reproduce the nuclear modification data at large proton-going rapidity and why the NLO pQCD calculations give so poor description of the individual $p+{\rm Pb}$ and $p+p$ spectra~\cite{Eskola:2019dui}. New measurements for this process, as presented by the CMS collaboration at this conference with preliminary Run~2 data, will be most helpful for addressing these issues. Notably, no direct measurement of the differential dijet cross section or its nuclear modification ratio have been made thus far. Only self-normalised spectra~\cite{CMS:2018jpl} and conditional-yield~\cite{ATLAS:2019jgo} measurements are currently available. It should be noted that while these kind of self-normalised quantities are helpful for cancelling systematic uncertainties and NP effects, they can also induce anti-correlation between data points even for statistical uncertainties~\cite{Eskola:2022rlm}.

\begin{figure}
\centering
\sidecaption
\includegraphics[width=6.6cm]{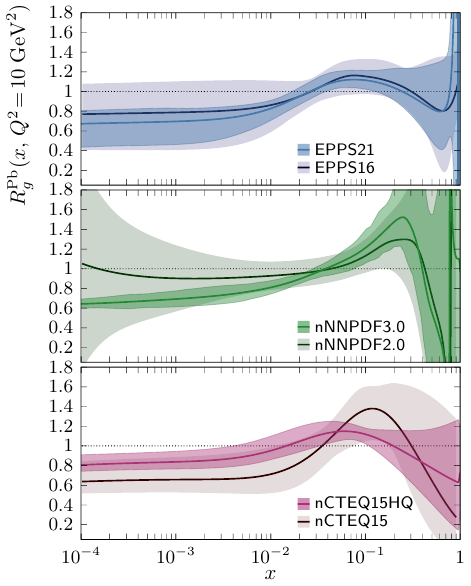}
\caption{A comparison of the nuclear modification factor $R_g^{\rm Pb} = f_g^{\rm Pb} / f_g^p$ of the gluon PDF in the lead nucleus as given by the EPPS21~\cite{Eskola:2021nhw}, nNNPDF3.0~\cite{AbdulKhalek:2022fyi} and nCTEQ15HQ~\cite{Duwentaster:2022kpv} analyses and their earlier versions.}
\label{fig-gluon-modification}
\end{figure}

Figure~\ref{fig-gluon-modification} compares the gluon nuclear modification factor from recent analyses~\cite{Duwentaster:2022kpv,AbdulKhalek:2022fyi,Eskola:2021nhw} and their earlier versions. All major global nPDF fits find significant reduction in gluon uncertainties when including LHC data. These constraints are driven by the dijets and HF data, but also EW bosons and light-hadron production provide complementary constraints.

\section{Inclusive UPC processes as novel probes}
\label{sec-upc}

\begin{figure}[htb]
\centering
\includegraphics[width=13.2cm]{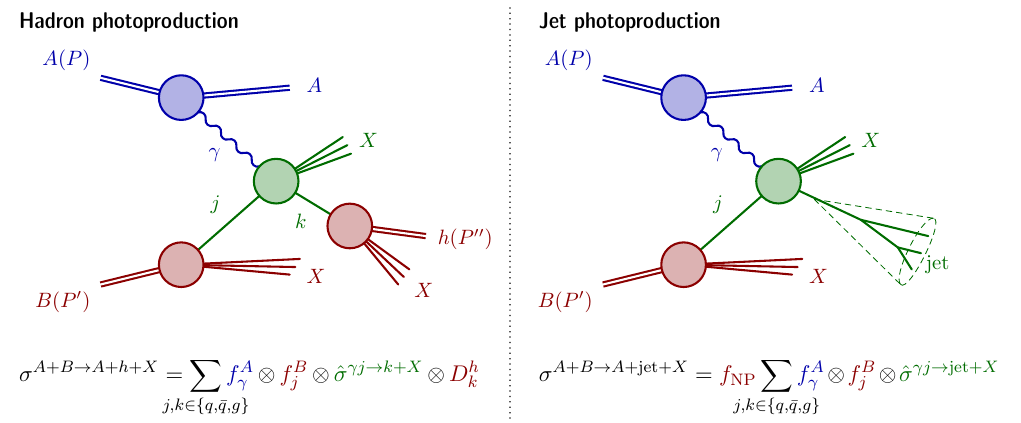}
\caption{Illustrations and pQCD expressions for the inclusive hadron (left) and jet (right) photoproduction processes in UPCs.}
\label{fig-upc}
\end{figure}

A lot of progress has been recently made in using $A+A$ UPC processes as probes of the nuclear contents~\cite{Guzey:2023zzb}. In particular, a new class of \emph{inclusive} UPC photoproduction processes is emerging as novel probes. At this conference, the first measurement of UPC dijet photoproduction~\cite{Strikman:2005yv,Guzey:2018dlm} by ATLAS~\cite{ATLAS:2024mvt} and preliminary results for the inclusive HF photoproduction~\cite{Baron:1993nk,Klein:2002wm} from CMS~\cite{CMS:2024ayq} and ALICE collaborations were presented. These processes are illustrated in figure~\ref{fig-upc}. The same hadronisation-treatment considerations have to be made as for the corresponding hadroproduction processes, but the underlying-event contribution can be expected to be smaller, making the UPCs a ``cleaner'' environment for measuring these final states. On the other hand, new complications arrise from the experimental selection of the UPC events. To identify these inclusive UPC processes and the direction of the photon-emitting nucleus, the experiments use a $0nXn$ forward-neutron event-class requirement together with associated rapidity-cut criteria. Correspondingly, theory predictions have to account for the electromagnetic dissociation (EMD) probability of the photon emitter~\cite{Baltz:2002pp} and for the fact that the diffractive contribution to the photo-nuclear cross section is excluded by this selection~\cite{Guzey:2020ehb}. The impact-parameter dependent EMD and hadronic-interaction probabilities enter the calculation of the \emph{effective} photon flux $f_\gamma^A(y)$ describing the distribution of quasi-real photons carrying a fraction $y$ of the energy of the photon-emitting nucleus, conditional to the UPC criteria. For processes requiring high photon energies this can even lead to sensitivity to the full trasverse-plane event geometry and the spatial structure of the target nucleus~\cite{Eskola:2024fhf}. Furthermore, the PDFs of real photons are currently somewhat poorly determined, which causes an uncertainty for the size of the resolved contribution to these processes~\cite{Helenius:2018mhx}. These matters have to be properly understood and the associated uncertainties quantified for the inclusive UPC processes to yield reliable information on the nPDFs.

\section{Outlook}
\label{summary}
Nuclear PDFs are being constrained by an increasing amount of LHC data with recent global fits including few thousand data points on a variety of hard and EW processes, showing that collinear factorisation works in $l+A$ and $h+A$ collisions across a large phase space. Still, significant differences appear e.g.\ between the gluon nPDF extractions, and studies with extended datasets including the LHC Run~2 HF data~\cite{LHCb:2019avm,LHCb:2022dmh} and the upcoming dijet measurements are needed in order to shed light on the observed tensions and further constrain the nPDFs to find limits where proposed additional effects (nonlinear saturation dynamics, cold nuclear matter energy loss, hadronisation nonuniversality) become significant for hard-probes phenomenology, or to outrule their impact. This requires precision in both data and pQCD calculations. Additional processes for these studies involve direct photons~\cite{ATLAS:2019ery,ALICE:2025bnc} and top-quark production~\cite{CMS:2017hnw,ATLAS:2024qdu} where the data agree with nPDF predictions, but the current experimental uncertainties are too large to yield stringent constraints. As an exciting development, inclusive UPC processes are now rapidly emerging as new nPDF probes.

\medskip
\begin{acknowledgement}%
The work of P.P.\ has been funded through the Research Council of Finland (projects 330448 and 331545) and as a part of the Center of Excellence in Quark Matter of the Research Council of Finland (project 364194).
\end{acknowledgement}


\begin{thebibliography}{88}
%
\bibitem{Akiba:2015vaa}
Y.~Akiba,
Quest for the quark--gluon plasma---hard and electromagnetic probes.
Prog.\ Theor.\ Exp.\ Phys.\ \textbf{2015}, 03A105 (2015).
\url{https://doi.org/10.1093/ptep/ptu080}
%
\bibitem{Apolinario:2022vzg}
L.~Apolin\'ario, Y.~J.~Lee and M.~Winn,
Heavy quarks and jets as probes of the QGP.
Prog.\ Part.\ Nucl.\ Phys.\ \textbf{127}, 103990 (2022).
\url{https://doi.org/10.1016/j.ppnp.2022.103990}
%
\bibitem{Pablos:2024yxw}
D.~Pablos,
Jet modifications and medium response - Theoretical overview.
EPJ Web Conf.\ \textbf{296}, 01028 (2024).
\url{https://doi.org/10.1051/epjconf/202429601028}
%
\bibitem{Caucal:2020uic}
P.~Caucal, E.~Iancu and G.~Soyez,
Jet radiation in a longitudinally expanding medium.
JHEP \textbf{04}, 209 (2021).
\url{https://doi.org/10.1007/JHEP04(2021)209}
%
\bibitem{Avramescu:2024poa}
D.~Avramescu, V.~Greco, T.~Lappi, H.~M\"antysaari and D.~M\"uller,
The impact of glasma on heavy flavor azimuthal correlations and spectra.
\url{https://doi.org/10.48550/arXiv.2409.10564}
%
\bibitem{Brewer:2021tyv}
J.~Brewer, A.~Huss, A.~Mazeliauskas and W.~van der Schee,
Ratios of jet and hadron spectra at LHC energies: Measuring high-$p_T$ suppression without a pp reference.
Phys.\ Rev.\ D \textbf{105}, 074040 (2022).
\url{https://doi.org/10.1103/PhysRevD.105.074040}
%
\bibitem{Paakkinen:2021jjp}
P.~Paakkinen,
Light-nuclei gluons from dijet production in proton-oxygen collisions.
Phys.\ Rev.\ D \textbf{105}, L031504 (2022).
\url{https://doi.org/10.1103/PhysRevD.105.L031504}
%
\bibitem{Gebhard:2024flv}
J.~Gebhard, A.~Mazeliauskas and A.~Takacs,
No-quenching baseline for energy loss signals in oxygen-oxygen collisions.
JHEP \textbf{04}, 034 (2025).
\url{https://doi.org/10.1007/JHEP04(2025)034}
%
\bibitem{Grosse-Oetringhaus:2024bwr}
J.~F.~Grosse-Oetringhaus and U.~A.~Wiedemann,
A Decade of Collectivity in Small Systems.
\url{https://cds.cern.ch/record/2904261}
%
\bibitem{Arleo:2021bpv}
F.~Arleo, G.~Jackson and S.~Peign\'e,
Impact of fully coherent energy loss on heavy meson production in pA collisions.
JHEP \textbf{01}, 164 (2022).
\url{https://doi.org/10.1007/JHEP01(2022)164}
%
\bibitem{Klasen:2023uqj}
M.~Klasen and H.~Paukkunen,
Nuclear PDFs After the First Decade of LHC Data.
Ann.\ Rev.\ Nucl.\ Part.\ Sci.\ \textbf{74}, 49-87 (2024).
\url{https://doi.org/10.1146/annurev-nucl-102122-022747}
%
\bibitem{Muzakka:2022wey}
K.~F.~Muzakka \textit{et al.},
Compatibility of neutrino DIS data and its impact on nuclear parton distribution functions.
Phys.\ Rev.\ D \textbf{106}, 074004 (2022).
\url{https://doi.org/10.1103/PhysRevD.106.074004}
%
\bibitem{Helenius:2024fow}
I.~Helenius, H.~Paukkunen and S.~Yrj\"anheikki,
Dimuons from neutrino-nucleus collisions in the semi-inclusive DIS approach.
JHEP \textbf{09}, 043 (2024).
\url{https://doi.org/10.1007/JHEP09(2024)043}
%
\bibitem{Paukkunen:2020rnb}
H.~Paukkunen and P.~Zurita,
Can we fit nuclear PDFs with the high-x CLAS data?.
Eur.\ Phys.\ J.\ C \textbf{80}, 381 (2020).
\url{https://doi.org/10.1140/epjc/s10052-020-7971-1}
%
\bibitem{Segarra:2020gtj}
E.~P.~Segarra \textit{et al.}
Extending nuclear PDF analyses into the high-$x$ , low-$Q^2$ region.
Phys.\ Rev.\ D \textbf{103}, 114015 (2021).
\url{https://doi.org/10.1103/PhysRevD.103.114015}
%
\bibitem{Accardi:2012qut}
A.~Accardi \textit{et al.}
Electron Ion Collider: The Next QCD Frontier: Understanding the glue that binds us all.
Eur.\ Phys.\ J.\ A \textbf{52}, 268 (2016).
\url{https://doi.org/10.1140/epja/i2016-16268-9}
%
\bibitem{Paukkunen:2010qg}
H.~Paukkunen and C.~A.~Salgado,
Constraints for the nuclear parton distributions from Z and W production at the LHC.
JHEP \textbf{03}, 071 (2011).
\url{https://doi.org/10.1007/JHEP03(2011)071}
%
\bibitem{Eskola:2022rlm}
K.~J.~Eskola, P.~Paakkinen, H.~Paukkunen and C.~A.~Salgado,
Proton-PDF uncertainties in extracting nuclear PDFs from $W^\pm $ production in p+Pb collisions.
Eur.\ Phys.\ J.\ C \textbf{82}, 271 (2022).
\url{https://doi.org/10.1140/epjc/s10052-022-10179-2}
%
\bibitem{Helenius:2021tof}
I.~Helenius, M.~Walt and W.~Vogelsang,
NNLO nuclear parton distribution functions with electroweak-boson production data from the LHC.
Phys.\ Rev.\ D \textbf{105}, 094031 (2022).
\url{https://doi.org/10.1103/PhysRevD.105.094031}
%
\bibitem{Duwentaster:2022kpv}
P.~Duwent\"aster \textit{et al.},
Impact of heavy quark and quarkonium data on nuclear gluon PDFs.
Phys.\ Rev.\ D \textbf{105}, 114043 (2022).
\url{https://doi.org/10.1103/PhysRevD.105.114043}
%
\bibitem{AbdulKhalek:2022fyi}
R.~Abdul Khalek, R.~Gauld, T.~Giani, E.~R.~Nocera, T.~R.~Rabemananjara and J.~Rojo,
nNNPDF3.0: evidence for a modified partonic structure in heavy nuclei.
Eur.\ Phys.\ J.\ C \textbf{82}, 507 (2022).
\url{https://doi.org/10.1140/epjc/s10052-022-10417-7}
%
\bibitem{Eskola:2021nhw}
K.~J.~Eskola, P.~Paakkinen, H.~Paukkunen and C.~A.~Salgado,
EPPS21: a global QCD analysis of nuclear PDFs.
Eur.\ Phys.\ J.\ C \textbf{82}, 413 (2022).
\url{https://doi.org/10.1140/epjc/s10052-022-10359-0}
%
\bibitem{ALICE:2022cxs}
S.~Acharya \textit{et al.},
W$^\pm$-boson production in p$-$Pb collisions at $\sqrt{s_{NN}} = 8.16$ TeV and PbPb collisions at $\sqrt{s_{NN}} = 5.02$ TeV.
JHEP \textbf{05}, 036 (2023).
\url{https://doi.org/10.1007/JHEP05(2023)036}
%
\bibitem{CMS:2021ynu}
A.~M.~Sirunyan \textit{et al.},
Study of Drell-Yan dimuon production in proton-lead collisions at $\sqrt{s_\mathrm{NN}} =$ 8.16 TeV.
JHEP \textbf{05}, 182 (2021).
\url{https://doi.org/10.1007/JHEP05(2021)182}
%
\bibitem{Duwentaster:2021ioo}
P.~Duwent\"aster \textit{et al.},
Impact of inclusive hadron production data on nuclear gluon PDFs.
Phys.\ Rev.\ D \textbf{104}, 094005 (2021).
\url{https://doi.org/10.1103/PhysRevD.104.094005}
%
\bibitem{LHCb:2022tjh}
R.~Aaij \textit{et al.},
Nuclear Modification Factor of Neutral Pions in the Forward and Backward Regions in p-Pb Collisions.
Phys.\ Rev.\ Lett.\ \textbf{131}, 042302 (2023).
\url{https://doi.org/10.1103/PhysRevLett.131.042302}
%
\bibitem{LHCb:2017yua}
R.~Aaij \textit{et al.},
Study of prompt D$^{0}$ meson production in $p$Pb collisions at $ \sqrt{s_{\mathrm{NN}}}=5 $ TeV.
JHEP \textbf{10}, 090 (2017).
\url{https://doi.org/10.1007/JHEP10(2017)090}
%
\bibitem{ALICE:2021est}
S.~Acharya \textit{et al.},
Nuclear modification factor of light neutral-meson spectra up to high transverse momentum in p\textendash{}Pb collisions at sNN=8.16 TeV.
Phys.\ Lett.\ B \textbf{827}, 136943 (2022).
\url{https://doi.org/10.1016/j.physletb.2022.136943}
%
\bibitem{Paakkinen:2022qxn}
P.~Paakkinen,
Nuclear PDFs at the beginning of LHC Run 3.
PoS \textbf{LHCP2022}, 137 (2023).
\url{https://doi.org/10.22323/1.422.0137}
%
\bibitem{CMS:2018jpl}
A.~M.~Sirunyan \textit{et al.},
Constraining gluon distributions in nuclei using dijets in proton-proton and proton-lead collisions at $\sqrt{s_{_\mathrm{NN}}} =$ 5.02 TeV.
Phys.\ Rev.\ Lett.\ \textbf{121}, 062002 (2018).
\url{https://doi.org/10.1103/PhysRevLett.121.062002}
%
\bibitem{Eskola:2019dui}
K.~J.~Eskola, P.~Paakkinen and H.~Paukkunen,
Non-quadratic improved Hessian PDF reweighting and application to CMS dijet measurements at 5.02 TeV.
Eur.\ Phys.\ J.\ C \textbf{79}, 511 (2019).
\url{https://doi.org/10.1140/epjc/s10052-019-6982-2}
%
\bibitem{ATLAS:2019jgo}
M.~Aaboud \textit{et al.},
Dijet azimuthal correlations and conditional yields in pp and p+Pb collisions at sNN=5.02TeV with the ATLAS detector.
Phys.\ Rev.\ C \textbf{100}, 034903 (2019).
\url{https://doi.org/10.1103/PhysRevC.100.034903}
%
\bibitem{Guzey:2023zzb}
V.~Guzey,
UPCs as probes of partonic structure \textendash{}exclusive and inclusive processes.
PoS \textbf{HardProbes2023}, 021 (2024).
\url{https://doi.org/10.22323/1.438.0021}
%
\bibitem{Strikman:2005yv}
M.~Strikman, R.~Vogt and S.~N.~White,
Probing small x parton densities in ultraperipheral AA and pA collisions at the LHC.
Phys.\ Rev.\ Lett.\ \textbf{96}, 082001 (2006).
\url{https://doi.org/10.1103/PhysRevLett.96.082001}
%
\bibitem{Guzey:2018dlm}
V.~Guzey and M.~Klasen,
Inclusive dijet photoproduction in ultraperipheral heavy ion collisions at the CERN Large Hadron Collider in next-to-leading order QCD.
Phys.\ Rev.\ C \textbf{99}, 065202 (2019).
\url{https://doi.org/10.1103/PhysRevC.99.065202}
%
\bibitem{ATLAS:2024mvt}
G.~Aad \textit{et al.},
Measurement of photonuclear jet production in ultraperipheral Pb+Pb collisions at sNN=5.02\,\,TeV with the ATLAS detector.
Phys.\ Rev.\ D \textbf{111}, 052006 (2025).
\url{https://doi.org/10.1103/PhysRevD.111.052006}
%
\bibitem{Baron:1993nk}
N.~Baron and G.~Baur,
Photon - hadron interactions in relativistic heavy ion collisions.
Phys.\ Rev.\ C \textbf{48}, 1999-2010 (1993).
\url{https://doi.org/10.1103/PhysRevC.48.1999}
%
\bibitem{Klein:2002wm}
S.~R.~Klein, J.~Nystrand and R.~Vogt,
Heavy quark photoproduction in ultraperipheral heavy ion collisions.
Phys.\ Rev.\ C \textbf{66}, 044906 (2002).
\url{https://doi.org/10.1103/PhysRevC.66.044906}
%
\bibitem{CMS:2024ayq}
CMS~Collaboration,
Constraining nuclear parton dynamics with the first measurement of D0-photoproduction in ultraperipheral heavy-ion collisions at the LHC.
CMS-PAS-HIN-24-003.
\url{https://cds.cern.ch/record/2910905}
%
\bibitem{Baltz:2002pp}
A.~J.~Baltz, S.~R.~Klein and J.~Nystrand,
Coherent vector meson photoproduction with nuclear breakup in relativistic heavy ion collisions.
Phys.\ Rev.\ Lett.\ \textbf{89}, 012301 (2002).
\url{https://doi.org/10.1103/PhysRevLett.89.012301}
%
\bibitem{Guzey:2020ehb}
V.~Guzey and M.~Klasen,
How large is the diffractive contribution to inclusive dijet photoproduction in ultraperipheral collisions at the LHC?.
Phys.\ Rev.\ D \textbf{104}, 114013 (2021).
\url{https://doi.org/10.1103/PhysRevD.104.114013}
%
\bibitem{Eskola:2024fhf}
K.~J.~Eskola, V.~Guzey, I.~Helenius, P.~Paakkinen and H.~Paukkunen,
Spatial resolution of dijet photoproduction in near-encounter ultraperipheral nuclear collisions.
Phys.\ Rev.\ C \textbf{110}, 054906 (2024).
\url{https://doi.org/10.1103/PhysRevC.110.054906}
%
\bibitem{Helenius:2018mhx}
I.~Helenius,
Probing nuclear PDFs with dijets in ultra-peripheral Pb+Pb collisions.
PoS \textbf{HardProbes2018}, 118 (2018).
\url{https://doi.org/10.22323/1.345.0118}
%
\bibitem{LHCb:2019avm}
R.~Aaij \textit{et al.},
Measurement of $B^+$, $B^0$ and $\Lambda_b^0$ production in $p\mkern 1mu\mathrm{Pb}$ collisions at $\sqrt{s_\mathrm{NN}}=8.16\,{\rm TeV}$.
Phys.\ Rev.\ D \textbf{99}, 052011 (2019).
\url{https://doi.org/10.1103/PhysRevD.99.052011}
%
\bibitem{LHCb:2022dmh}
R.~Aaij \textit{et al.},
Measurement of the Prompt D0 Nuclear Modification Factor in p-Pb Collisions at sNN=8.16\,\,TeV.
Phys.\ Rev.\ Lett.\ \textbf{131}, 102301 (2023).
\url{https://doi.org/10.1103/PhysRevLett.131.102301}
%
\bibitem{ATLAS:2019ery}
M.~Aaboud \textit{et al.},
Measurement of prompt photon production in $\sqrt{s_\mathrm{NN}} = 8.16$ TeV $p$+Pb collisions with ATLAS.
Phys.\ Lett.\ B \textbf{796}, 230-252 (2019).
\url{https://doi.org/10.1016/j.physletb.2019.07.031}
%
\bibitem{ALICE:2025bnc}
S.~Acharya \textit{et al.},
Measurement of isolated prompt photon production in pp and p-Pb collisions at the LHC.
\url{https://doi.org/10.48550/arXiv.2502.18054}
%
\bibitem{CMS:2017hnw}
A.~M.~Sirunyan \textit{et al.},
Observation of top quark production in proton-nucleus collisions.
Phys.\ Rev.\ Lett.\ \textbf{119}, 242001 (2017).
\url{https://doi.org/10.1103/PhysRevLett.119.242001}
%
\bibitem{ATLAS:2024qdu}
G.~Aad \textit{et al.},
Observation of $ t\overline{t} $ production in the lepton+jets and dilepton channels in p+Pb collisions at $ \sqrt{s_{\textrm{NN}}} $ = 8.16 TeV with the ATLAS detector.
JHEP \textbf{11}, 101 (2024).
\url{https://doi.org/10.1007/JHEP11(2024)101}
%
\end{thebibliography}
\end{document}